\documentclass[10pt,5p,times,twocolumn]{elsarticle}

 \usepackage{graphicx}
\usepackage{color}

\usepackage{amssymb}
\usepackage{amsmath}

\begin{document}

\begin{frontmatter}

\title{Specific heat of a non-local attractive Hubbard model}

\author[ufsm]{E. J. Calegari\corref{cor1}}
\cortext[cor1]{eleonir@ufsm.br}
\author[ufsm]{C. O. Lobo}

\address[ufsm]{Laborat\'orio de Teoria da Mat\'eria Condensada,
Departamento de F\'{\i}sica - UFSM, 97105-900, Santa Maria, RS, Brazil}

\author[uff]{S. G. Magalhaes}
\address[uff]{Instituto de F\'isica, Universidade Federal Fluminense Av. Litor\^anea s/n, 24210,
346, Niter\'oi, Rio de Janeiro, Brazil}

\author[cbpf]{C. M. Chaves}
\author[cbpf]{ A. Troper}

\address[cbpf]{Centro Brasileiro de Pesquisas F\'{\i}sicas, Rua Xavier Sigaud 150, 22290-180,
 Rio de Janeiro, RJ, Brazil}

\begin{abstract}
The specific heat 
of an attractive (interaction $G<0$) non-local Hubbard model 
is investigated. We use a two-pole approximation which leads to a set of correlation functions. In particular, the correlation function 
$\langle\vec{S}_i\cdot\vec{S}_j\rangle$ plays an important role as a source of anomalies in the normal state of the model. 
Our results show that for a giving range of $G$  and $\delta$ where $\delta=1-n_T$
($n_T=n_{\uparrow}+n_{\downarrow}$), the specific heat as a function of the temperature 
presents a two peak structure.
Nevertehelesss, the presence of a pseudogap on the anti-nodal points $(0,\pm\pi)$ and $(\pm\pi,0)$ eliminates the two peak structure, the low temperature peak remaining.
The effects of the second nearest neighbor hopping on the specific heat are also investigated.
 
\end{abstract}

\begin{keyword}
 D. Hubbard model \sep D. pseudogap \sep D. specific heat 



\end{keyword}

\end{frontmatter}



\section{Introduction}
The phenomenology of high-$T_{c}$ Superconductors (HTSC) has brought 
several fundamental issues \cite{Taillefer}. 
One of these issues is certainly the 
nature of the pseudogap found in some of those materials.
In the possible competing scenarios 
on the nature of the pseudogap, one should mention two of them. Assuming that the pseudogap occurs below a temperature $T^{*}$: 
(i) the pseudogap would be due to the formation of incoherent pairs until that a superconducting phase develops below the critical temperature $T_c$ \cite{Randeria,Takami}; 
(ii) the pseudogap would be due to short-range fluctuations of magnetic 
nature which below a certain temperature $T_{ho}$ would give rise an ordered state 
ending at a Quantum Critical Point (QCP) which can coexist with the SC phase
\cite{Taillefer2}. Nevertheless, despite of the intense 
debate,  the complete explanation for the nature of the pseudogap is clearly an unsolved question.

Quite recently, an attractive Hubbard model with non-local interaction \cite{Calegari,Calegari1}
has been considered using a two pole approximation \cite{Roth,Edwards}. Although our model is not fully realistic for HTSC, 
it allows superconductivity with $d_{x^2-y^2}-$wave symmetry \cite{dagotto} and also can be quite useful to bring information on the possible sources of the pseudogap.
In Refs. \cite{Calegari,Calegari1}, it has been obtained the evolution of the Fermi surface 
from a closed shape to a hole pocket shape as well as the behavior of the $\chi$ with a maximum at $\delta^{*}$ where $\delta=1-n_T$ ($n_T=n_{\uparrow}+n_{\downarrow}$), and then, decreasing when 
$\delta<\delta^{*}$.
Remarkably, both results can be traced from one single mechanism, i. e., from short range antiferromagnetic (AF) correlations. To be precise, for a proper range of 
temperature and doping these correlations distort the renormalized quasi-particles bands shifting by $\Delta\epsilon$ the flat region on the anti-nodal
points ($(0,\pm\pi)$ and $(\pm\pi,0)$) to energies below the chemical potential \cite{Calegari1}. As a consequence, a pseudogap $\Delta_{PG} \approx\Delta\epsilon$, emerges
in the DOS close to the chemical potential. 

The mechanism discussed in the previous paragraph
also favors the existence of $d$-wave superconductivity
in the attractive non-local Hubbard model \cite{Calegari}. This occurs because the magnetic correlations 
enhance the density of states at the van Hove singularity (VHS) providing more electrons able to form superconducting pairs.
It should be noticed, that is exactly the non-locality of the attractive interacting term which triggers 
the short range AF correlations within the two pole approximation  for the attractive Hubbard model.

The electronic specific heat $C(T)$ is an important quantity giving relevant information about the 
pseudogap and the mechanisms behind it. Mostly important, the close relation between the specific heat and the density of states
allows a theoretical investigation about effects of correlations, in particular, the magnetic ones. 

Therefore, we present here a systematic study of the specific heat using the attractive non-local Hubbard model within the two pole approximation.
This approximation allows to deal properly with the regime of strong correlations. In particular, we assume that there is also next-neighbor hopping. As  will be shown below, 
this hopping furnishes an additional mechanism to amplify magnetic correlations and, therefore modifies the DOS, thus
affecting the specific heat.

\section{The model}
 \label{model}
The model investigated is a two-dimensional one-band Hubbard model \cite{Calegari,amos1}
which is given by:
\begin{equation}
H=\sum_{\langle \langle ij \rangle\rangle \sigma} t_{ij}d_{i\sigma}^{\dag}d_{j\sigma} +\frac{G}{2}\sum_{\langle ij\rangle \sigma} n_{i,\sigma} n_{j,-\sigma} -\mu\sum_{i\sigma}n_{i\sigma}
\label{eqH1}
\end{equation}
where $d_{i\sigma }^{\dag }(d_{i\sigma })$ is the fermionic creation
(annihilation) operator at site $i$ with spin $\sigma
=\{\uparrow ,\downarrow \}$ and $n_{i,\sigma }=d_{i\sigma }^{\dag
}d_{i\sigma }$ is the number operator. The quantity $t_{ij}$ represents the hopping between sites $i$ and $j$
and $\langle \langle ...\rangle \rangle $ indicates the sum over the first
and the second-nearest-neighbors of $i$ and $\mu $ is the chemical
potential. 
The second term in $H$ takes into account the  interaction between the electrons
in which $G$ is a non-local attractive potential. 
The bare dispersion relation is given by 
%
${\varepsilon }_{\vec{k}}=2t[\cos (k_{x}a) +\cos (
k_{y}a)] +4t_{2}\cos ( k_{x}a) \cos (k_{y}a)$ 
%
where $t$ is the first-neighbor and $t_{2}$ is the second-neighbor hopping
amplitudes and $a$ is the lattice parameter.

In the two-poles approximation proposed by Roth \cite{Roth,Edwards}, the Green's function matrix is
defined as 
%
$\mathbf{{\cal {G}}}\left( \omega \right) =\mathbf{N}\left( \omega \mathbf{N-E}\right)
^{-1}\mathbf{N}$
%
in which $\mathbf{N}$ and $\mathbf{E}$ are the normalization and the energy
matrices, respectively \cite{Edwards}. 
\begin{figure}[t]                     
\centering
\includegraphics[angle=-90,width=9cm]{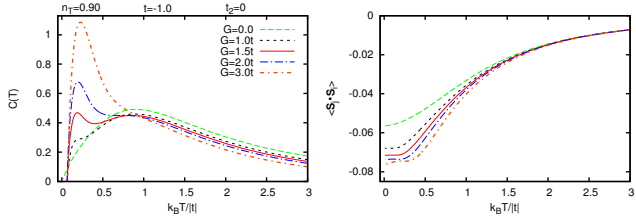}
\caption{The left panel shows the specific heat as a function of the temperature for different values of $G$. The two peaks structure of $C(T)$ 
is clear for $G=1.5t$. The right panel shows the spin-spin correlation function.}
\label{fig111}
\end{figure}
\begin{figure*}[t]
\centering                            
\includegraphics[angle=-90,width=15cm]{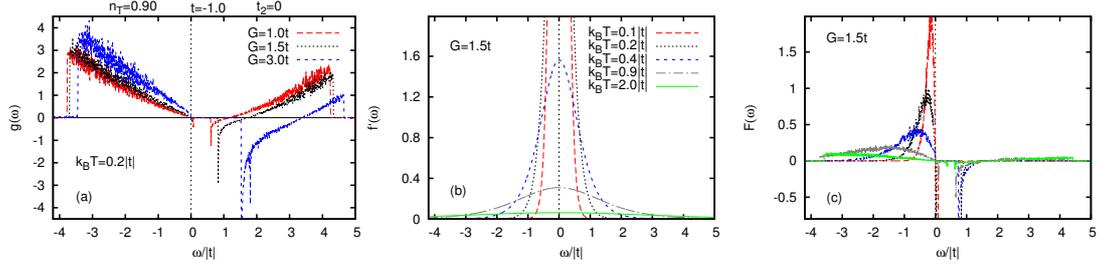}
\caption{In (a), the function $g(\omega)$ for different $G$ values. In (b), the function $f'(\omega)$ for different temperatures.
In (c), the function $F(\omega)$ for the same parameters as in (b). The parameters $n_T=0.90$, $t=-1.0$ and $t_2=0.0$, are common
for the figures (a), (b) and (c).}
\label{gwfwFwG}
\end{figure*}
%


The specific heat is given by 
$C(T)=\frac{\partial {E}}{\partial {T}}$
%
where ${E}$ is the energy per atom \ and ${T}$ the temperature.
In the grand canonical ensemble,
the energy is a function of the chemical potential $E\equiv E\left(\mu (T)\right)$,
where $\mu$ changes with the temperature. Therefore, the calculation of $C(T)$ must be performed keeping $\langle n\rangle$ 
constant in the $T-\mu$ plane \cite{moreo}.   
The energy per atom is $E=\frac{\langle H\rangle}{N}$ ($N$ being the number of sites of the system) and
can be written as \cite{Kishore}:
\begin{equation}
E=\frac{i}{2N}\lim_{\delta \rightarrow
0^{+}}\sum_{{\vec{k}},\sigma }\int_{-\infty }^{\infty }f(\omega)\ ({
\omega }+{\mu }+{\varepsilon }_{\vec{k}})
[{\cal {G}}_{{{\vec{k}},\sigma }}({\omega }+i{
\delta })-{\cal {G}}_{{{\vec{k}},\sigma }}({\omega }-i{\delta })]d{\omega }
\label{eqE}
\end{equation}
where $f(\omega)$ is the Fermi function and  ${\cal {G}}_{{{\vec{k}},\sigma }}({\omega })$ is a Green's function of the type
\begin{equation}
{\cal {G}}_{{{\vec{k}},\sigma }}({\omega })=\frac{Z_{1,\sigma }({{\vec{k}}})}{\omega
-\omega _{1,\sigma }(\vec{k})}+\frac{Z_{2,\sigma }(\vec{k})}{\omega -\omega
_{2,\sigma }(\vec{k})} 
\label{eqG}
\end{equation}
with the spectral weights $Z_{i,\sigma }(\vec{k})$ and the renormalized bands $\omega
_{i,\sigma }(\vec{k})$ defined in App ( \ref{app:A}).

Combining $C(T)=\frac{\partial {E}}{\partial {T}}$ with equations
(\ref{eqE}) and (\ref{eqG}) we obtain
%
$C(T)=\int_{-\infty }^{\infty }F(\omega)d{\omega }$
%
with 
%
$F(\omega)=f'(\omega )g(\omega )$
%
and $f'(\omega)=\frac{1}{\omega}\frac{\partial {f({\omega })}}{\partial {T}}$.
The function $g(\omega )$ is defined as
\begin{equation}
g(\omega )=\frac{1}{2N}\sum_{i=1}^{2}\sum_{{\vec{k}},\sigma }\widetilde{Z}_{i,\sigma }(\vec{k})\delta (\omega -\omega
_{i,\sigma }(\vec{k})) 
\label{B34}
\end{equation}
where
$\widetilde{Z}_{i,\sigma }(\vec{k})=\left(\omega _{i,\sigma }(\vec{k})+{\mu }+{\varepsilon }_{\vec{k}}\right)
\omega_{i,\sigma }(\vec{k}) Z_{i,\sigma }(\vec{k})$.

 The spin-spin correlation function discussed in the numerical results section is given by:
$\langle \vec{S_j}\cdot\vec{S_i}\rangle= \langle S_j^zS_i^z \rangle - h_{ij,-\sigma}$
with $\langle S_j^zS_0^z \rangle$ and $h_{ij,-\sigma}$ defined in App ( \ref{app:A}).

\section{Numerical results}

\subsection*{(i) Specific heat for $t_2=0$}

The specific heat for the Hubbard model with only nearest neighbor hopping, is shown in the left panel in
figure \ref{fig111}, for $n_T=0.90$ and different values for the interaction $G$. 
For $G=0.0$, the specific heat  presents a  Schottky anomaly \cite{Stout} with a maximum at $k_BT\simeq 0.90|t|$. If $|G|$ is increased, the 
peak in $k_BT\simeq 0.90|t|$ is preserved and a second peak emerges in $k_BT\simeq 0.20|t|$, as can be observed for  $G=1.5t$. 
For high values of $|G|$, for instance $G=3.0t$, the peak at low temperature dominates and the specific heat shows only a single peak.
The right panel in figure  \ref{fig111} shows the spin-spin correlation function $\langle\vec{S}_i\cdot\vec{S}_j \rangle$.
Another feature observed in $C(T)$ is the presence of a crossing point in $k_BT\approx0.8|t|$. Such crossing point is a characteristic of
the specific heat of many correlated systems \cite{vollhardt}.
\begin{figure*}[t]
\centering
\includegraphics[angle=-90,width=14cm]{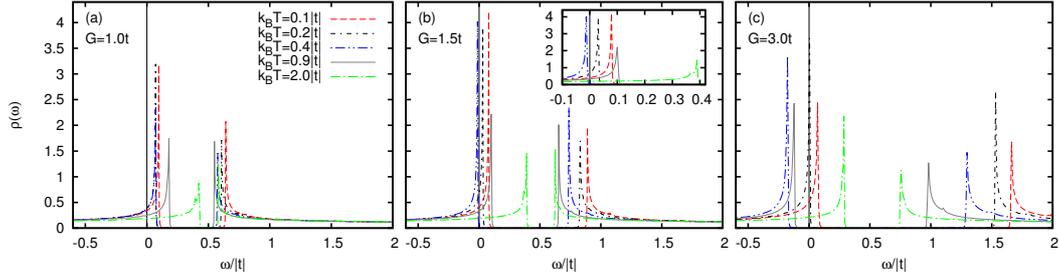}
\caption{The density of states for $n_T=0.90$, $t=-1.0$ and $t_2=0$. The vertical line in $\frac{\omega}{|t|}=0$ indicates the position of the chemical potential $\mu$. }
\label{dos}
\end{figure*}
\begin{figure}
\centering                              
\includegraphics[angle=-90,width=7cm]{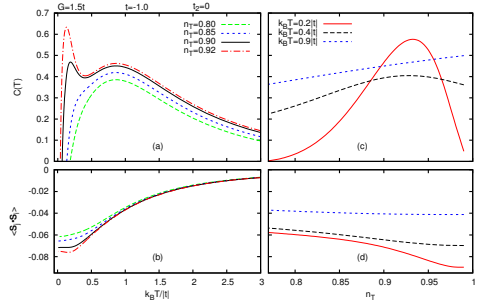}
\caption{In (a) and (b), the specific heat and $\langle\vec{S}_i\cdot\vec{S}_j \rangle$ 
as a function of temperature with different values of $n_T$. In (c) and (d) the specific heat and the $\langle\vec{S}_i\cdot\vec{S}_j \rangle$ 
as a function of $n_T$ and different temperatures. }
\label{cn}
\end{figure}

In order to understanding the temperature dependence of $C(T)$, let us analyze $F(\omega)$ 
and $g(\omega)$   defined  above.   
Figure \ref{gwfwFwG}(a) shows $g(\omega)$ for $n_T=0.90$ and different values of $G$. 
The figure \ref{gwfwFwG}(b) presents $f'(\omega)$ for $G=1.5t$ and different values of $k_BT$. 
The $F(\omega)$ is shown in figure \ref{gwfwFwG}(c).
The $g(\omega)$ shown in figure \ref{gwfwFwG}(a) presents an important feature, namely, both the positive area associated to the 
lower Hubbard band and the negative area related to the upper Hubbard band enhance as $|G|$ increases. As a consequence, the low temperature peak on $C(T)$
is favored by $G$ while the high temperature peak is suppressed by $G$, as can be observed in figure \ref{fig111}.
For intermediate values of $G$, there is a temperature in which the negative area in $g(\omega)$ associated to the upper Hubbard band gives the maximum
contribution to $F(\omega)$ leading to a local minimum in $C(T)$ at this temperature.
This is the reason why at moderated values of $G$, both the low and high temperature peaks  coexist on $C(T)$ as can be observed in figure \ref{fig111}, for $G=1.5t$.

The right panel in figure \ref{fig111} shows that $|\langle\vec{S}_i\cdot\vec{S}_j \rangle|$ is large at low temperatures
and increases with $|G|$. As discussed in references \cite{Calegari,Calegari1,Eleonir2011}, the $\langle\vec{S}_i\cdot\vec{S}_j \rangle$ modify the renormalized
band structure by enlarging the flat region near the anti-nodal points $(\pi,0)$ and $(0,\pi)$. 
As a consequence, a pronounced peak emerges on the density of state. If such peak is near the chemical potential it gives a strong 
contribution to the specific heat. At high temperatures  $|\langle\vec{S}_i\cdot\vec{S}_j \rangle|$ decreases and its effect becomes  negligible. 
The density of states for different temperatures and $G$ are shown in figure \ref{dos}. In \ref{dos}(a), $G=1.0t$ and the chemical potential $\mu$
intercepts the density of states $\rho(\omega)$ below the peak associated to a 
 VHS, for all values of temperatures shown in the figure.
In \ref{dos}(b), in which $G=1.5t$, $\rho(\omega=\mu)$ is maximum for $k_BT=0.2|t|$, while for $k_BT=0.4|t|$, $\mu$  intercepts $\rho(\omega)$ after the VHS 
where $\rho(\omega=\mu)\backsimeq 0$ (see the inset). It is interesting to note that the specific heat presents a peak just at $k_BT=0.2|t|$ and a local minimum 
at $k_BT=0.4|t|$ (see figure \ref{fig111}).  This occurs
because at low temperatures the function $f'(\omega)=\frac{1}{\omega}\frac{\partial {f({\omega })}}{\partial {T}}$
($f(\omega)$ is the Fermi function) is very close to the chemical potential. In this case the position of chemical
potential on $\rho(\omega)$ plays an important role. For $G=3.0t$, figure \ref{dos}(c) shows that the chemical potential intercepts $\rho(\omega)$ on the 
VHS when $k_BT=0.2|t|$. However, for $k_BT=0.4|t|$ and $k_BT=0.9|t|$, $\mu$ is found within the gap, 
where $\rho(\omega)=0$.
Figure \ref{cn}(a) shows the specific heat for different values of $n_T$. Notice that when $n_T$ decreases the peak at low temperature disappears.
This occurs because the $|\langle\vec{S}_i\cdot\vec{S}_j \rangle|$ becomes small decreasing the density of states at the VHS. Moreover,
the low occupation moves the chemical potential $\mu$  away from the VHS. 
On the other hand, if $n_T$ increases,  $|\langle\vec{S}_i\cdot\vec{S}_j \rangle|$ is enhanced, $\mu$ moves closer 
to the VHS and as consequence the low temperature peak in $C(T)$ enlarges. Figure \ref{cn}(b) shows 
the spin-spin correlation function $\langle\vec{S}_i\cdot\vec{S}_j \rangle$ for the same parameters as in the upper panel.
Figure \ref{cn}(c) shows that
at low temperatures, the specific heat exhibits a maximum for a given $n_T$. 
At high temperatures the maximum disappears. Figure \ref{cn}(d) displays the spin-spin correlation function $\langle\vec{S}_i\cdot\vec{S}_j \rangle$ for the same parameters as in the upper panel.

\begin{figure}
\centering
\includegraphics[angle=-90,width=9cm]{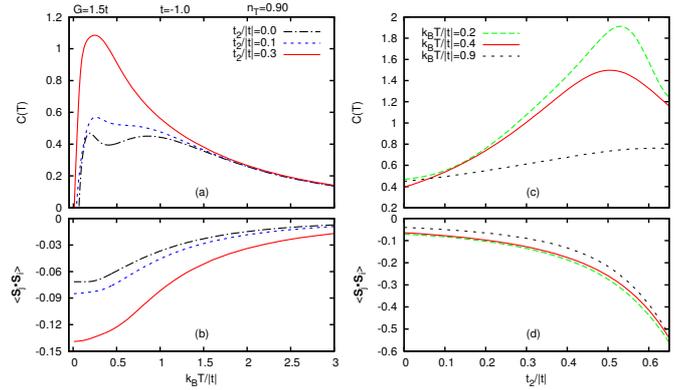}
\caption{ In (a) and (b) the specific heat spin-spin correlation function $\langle\vec{S}_i\cdot\vec{S}_j \rangle$ as a function of temperature with different values of $\frac{t_2}{|t|}$.
In (c) and (d) the specific heat and the $\langle\vec{S}_i\cdot\vec{S}_j \rangle$ as a function of  $t_2/|t|$ and three distinct temperatures.}
\label{ctl}
\end{figure} 
\begin{figure}
\centering
\includegraphics[angle=-90,width=9.5cm]{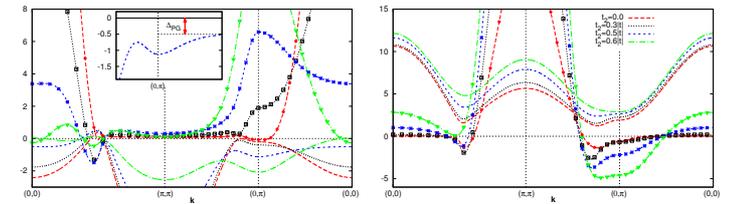}
\caption{ The lines with symbols show the effective spectral weight $\widetilde{Z}_{i,\sigma }(\vec{k})$ (see equation (\ref{B34})) and the lines with no symbols show 
the renormalized bands $\omega_{i,\sigma }(\vec{k})$. The model parameters are $t=-1.0$, $n_T=0.90$, $G=1.5t$ and the temperature is $k_BT/|t|=0.2$.}
\label{w1Z1}
\end{figure}
\subsection*{(ii) Specific heat for $t_2\neq0$}
The figure  \ref{ctl}(a) presents the specific heat as a function 
of temperature for different values of the second nearest-neighbors amplitude $t_2$. 
The low temperature peak in the specific heat is strongly enhanced by 
$t_2/|t|$.
This occurs because $t_2$ enhances $\langle\vec{S}_i\cdot\vec{S}_j \rangle$ and enlarges the flat regions on the renormalized bands resulting 
in a high density of states on the  VHS.
As a consequence, the low temperature peak on $C(T)$ increases while the high temperature peak is not affected
because temperature suppresses $\langle\vec{S}_i\cdot\vec{S}_j \rangle$. The lower panel in figure  \ref{ctl}(b) shows the behavior of  $\langle\vec{S}_i\cdot\vec{S}_j \rangle$ for the same parameters as in \ref{ctl}(a).
Figure \ref{ctl}(c) shows 
that the effects of $t_2/|t|$ on $C(T)$ are more intensive at low temperatures
where $\langle\vec{S}_i\cdot\vec{S}_j \rangle$ is stronger. Furthermore, there is a maximum on $C(T)$ in $t_2/|t|\approx 0.5$ but, this maximum does not show at high temperatures. Figure  \ref{ctl}(d) shows   $\langle\vec{S}_i\cdot\vec{S}_j \rangle$
for the same parameters has in \ref{ctl}(c).
In figure \ref{w1Z1} we present the renormalized band structures and the effective spectral weights for $k_BT/|t|=0.2$ and several values of 
$t_2/|t|$. We observe that a pseudogap $\Delta_{PG}$ emerges from $t_2/|t|\simeq 0.2$ and persists until  $t_2/|t|\simeq 0.6$. Moreover, the pseudogap is maximum 
for $t_2/|t|\simeq 0.5$ which is just the value of  $t_2/|t|$ for which $C(T)$ is maximum (see figure \ref{ctl}). The inset in the upper panel of figure \ref{w1Z1}
shows in detail the pseudogap $\Delta_{PG}$ for $t_2/|t|= 0.5$. Indeed, for $t_2/|t|= 0.5$, the presence of the pseudogap gives rise to a wide flattening in 
$\omega_{1,\sigma }(\vec{k})$
along the direction $(0,\pi)$-$(0,0)$, which increases the density of states at the VHS
and also produces a peak on $g(\omega)$.
When $t_2/|t|$ increases, the pseudogap opens and the region of $\widetilde{Z}_{1,\sigma }(\vec{k})$ near $(0,\pi)$
becomes positive resulting in an enhancement of the specific heat. Nevertheless, for  $t_2/|t|\gtrsim 0.5$, the pseudogap starts to decrease and closes for 
$t_2/|t|\sim 0.6$. The lower panel in figure \ref{w1Z1} shows that the negative region on $\widetilde{Z}_{2,\sigma }(\vec{k})$ increases with $t_2/|t|$. However, at low temperatures such regions do not contribute to $C(T)$ because they are associated to the upper Hubbard band. On the other hand, when 
the temperature increases, the negative regions of $\widetilde{Z}_{2,\sigma }(\vec{k})$ become relevant and affect the specific heat
(see $C(T)$ for $k_BT/|t|=0.9$ in the upper panel in figure \ref{ctl}(c)).  
\begin{figure}
\centering
\includegraphics[angle=-90,width=9cm]{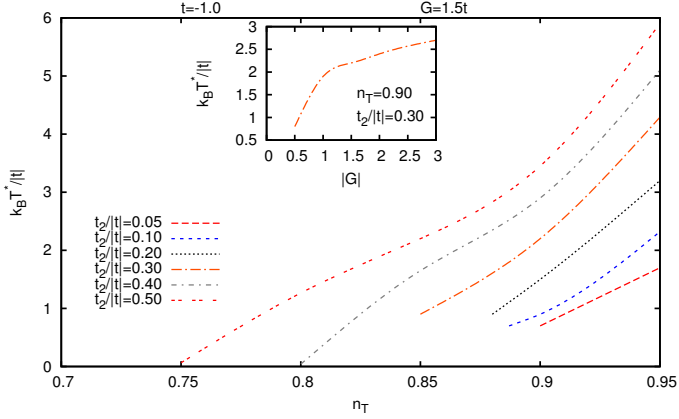}
\caption{ The temperature ($k_BT^*/|t|$) below which a pseudogap appears on the density of states. 
The main figure shows $k_BT^*/|t|$ as a function of the total occupation $n_T$ for different values of $t_2/|t|$. The inset shows 
$k_BT^*/|t|$ as a function of the modulus of the interaction $G$. }
\label{figPG}
\end{figure}

Figure \ref{figPG}  displays the temperature $T^*$ below which a pseudogap appears on the renormalized band as shown in figure  \ref{w1Z1} (left panel).
The pseudogap lines are displayed for different intensities $t_2/|t|$ and a common feature is that  $k_BT^*/|t|$ increases with the total occupation $n_T$.
Nevertheless, we observed that from $t_2/|t|=0.05$ to $t_2/|t|=0.30$ the pseudogap lines start in ($k_BT^*/|t|>0$). This occurs because above $n_T\simeq 0.80$ 
and below a critical value 
of $k_BT^*/|t|$, the spin-spin correlations become so strong that distort the renormalized band sufficiently to close the pseudogap. Therefore, in the present
scenario, the pseudogap is observed in a range of $G$, $t_2$, $n_T$ and $T$ in which the spin-spin correlations are typical for a ($G/t \geq 1$) regime.
The inset in figure \ref{figPG} shows the pseudogap line as a function of $|G|$. The $k_BT^*/|t|$ increases with $|G|$ because $G$ favors the correlations
$|\langle\vec{S}_i\cdot\vec{S}_j \rangle|$ (see the right panel of figure \ref{fig111}) which, in the present scenario, are responsible for the pseudogap.  
For $|G|\lesssim 0.5|t|$, the systems becomes weakly correlated and the pseudogap closes because the chemical potential reaches the upper Hubbard band.

\section{Conclusions}

The specific heat $C(T)$ of an attractive extended  Hubbard model has been studied within a two-pole approximation \cite{Roth,Edwards}. 
It has been verified that $C(T)$ as a function of temperature shows a two peak structure.
A systematic analysis of $C(T)$ in terms of the renormalized band structure allowed us to identify the mechanisms behind the two peak structure.
Indeed, the low temperature peak is associated to the lower Hubbard band while the high temperature peak is related to the upper Hubbard band. 
If $\mu$
is near a van Hove singularity (VHS) the low temperature peak is enhanced due to the close relation between specific heat and density of states.
On the other hand, the upper Hubbard band contributes with a positive and also a negative portion for $C(T)$.
For small $G$, the contribution is positive but when $|G|$ increases the negative portion dominates and the high temperature peak on $C(T)$ is suppressed. 
It should be stressed that the AF magnetic correlations associated to the spin-spin correlation function 
$\langle\vec{S}_i\cdot\vec{S}_j \rangle$ affect the behavior of the peaks on the specific heat. 
This occurs because  $\langle\vec{S}_i\cdot\vec{S}_j \rangle$ changes the renormalized band structure resulting in an enhancement or a decrease of the density of states, mainly, on the VHS. The lower temperature peak is more affected by this effect because the VHS contributes significantly to that peak which is the only one preserved when $|G|$ increases.
The low temperature peak is also deeply dependent on the occupation $n_T$. At low $n_T$ the chemical potential $\mu$ is far from the VHS.
Therefore, the low density of states on $\mu$ is not sufficient to induces the low temperature peak on $C(T)$. 

It has been verified that if the second nearest-neighbor hopping $t_2/|t|$ is present,
these same $\langle\vec{S}_i\cdot\vec{S}_j \rangle$ correlations induce a pseudogap on the renormalized band structure.
The pseudogap opens at the anti-nodal points $(0,\pm\pi)$ and $(\pm\pi,0)$ suggesting a $d$-wave symmetry for it.
Nevertheless, the pseudogap and the two peak structure on $C(T)$ do not coexist. Indeed,
the second nearest-neighbor hopping $t_2/|t|$ enhances the spin-spin correlations which increases the low temperature peak on $C(T)$ and suppresses the high temperature peak. 

In summary, the present results for an attractive non-local Hubbard model
suggest that in presence of a pseudogap the specific heat has a single peak structure
which is closely related to short-range AF magnetic correlations.  

\appendix 

\section{}

\label{app:A}
Within the two-pole approximation proposed in reference \cite{Roth}, the spectral weights and renormalized bands 
have the general form:
\begin{equation}
 Z_{i,\sigma }(\vec{k})=\frac{1}{2}-(-1)^i\left[\frac{\alpha-\varepsilon_{\vec{k}}+W_{\vec{k},\sigma}}{2X_{\vec{k},\sigma}}\right]
\end{equation}

\begin{equation}
 \omega_{i,\sigma }(\vec{k})=\frac{\beta+\varepsilon_{\vec{k}}+W_{\vec{k},\sigma}-2\mu}{2}
 +(-1)^i\left(\frac{X_{\vec{k},\sigma}}{2}\right).
\end{equation} 

Also,
$\alpha= \frac{G_2+n_{-\sigma}(G_1-2G_2)}{n_{-\sigma}(1-n_{-\sigma})}-2G_1$,
$\beta= \frac{G_2+n_{-\sigma}(G_1-2G_2)}{n_{-\sigma}(1-n_{-\sigma})}$
and
$X_{\vec{k}\sigma}=\sqrt{(\overline{G}-\varepsilon_{\vec{k}}+W_{\vec{k}\sigma})^2+4G_1(\varepsilon_{\vec{k}}-W_{\vec{k}\sigma})+\widetilde{G}}$
with

$\overline{G}=\frac{G_2+n_{-\sigma}(G_1-2G_2)}{n_{-\sigma}(1-n_{-\sigma})}$
and
$\widetilde{G}=\frac{4G_2(G_2-G_1)}{n_{-\sigma}(1-n_{-\sigma})}$.
The effective interactions $G_{1}$ and $G_{2}$ are defined as 
$G_1=G\sum_l\langle n_{l,-\sigma}\rangle$ and $G_2=G\sum_l\langle n_{l,-\sigma}n_{i,-\sigma}\rangle$
where $\langle n_{i\sigma}n_{j\sigma}\rangle=n_{\sigma}^2-\frac{a_{ij\sigma}n_{ij\sigma}+b_{ij\sigma}m_{ij\sigma}}{1-b_{ii,\sigma}b_{ii,-\sigma}}$
with 
$a_{ij,-\sigma}=\frac{n_{ij,-\sigma}-m_{ij,-\sigma}}{1-n_{\sigma}}$ and $b_{ij,-\sigma}=\frac{m_{ij,-\sigma}-n_{ij,-\sigma}n_{\sigma}}{n_{\sigma}(1-n_{\sigma})}$. 
The band shift $W_{\vec{k}\sigma }$ is given by 
$W_{\vec{k}\sigma } =\frac{1}{n_{\sigma}(1-n_{\sigma})}\frac{1}{N}\sum_{\vec{q}}\epsilon(\vec{k}-\vec{q})
F_{\sigma}(\vec{q})$,
where $\epsilon(\vec{k}-\vec{q})=\sum_{\langle\langle i=0\rangle\rangle j\neq 0}t_{0j}e^{i(\vec{k}-\vec{q})\cdot \vec{R}_j}$
and $F_{\sigma}(\vec{q})$ is given in terms of $\langle \vec{S_j}\cdot\vec{S_i}\rangle$ and the Fourier transform of $n_{j0\sigma }$ and $m_{j\sigma}$ defined as $n_{0j\sigma}=\langle d_{0\sigma}^{\dagger}d_{j\sigma}\rangle=\frac{1}{N}
\sum_{\vec k}{\cal{F}}_{\omega}G_{\vec{k}\sigma}^{(11)}e^{i\vec{k}\cdot \vec{R}_j}$ and $m_{j\sigma}=\langle d_{0\sigma}^{\dagger}n_{j-\sigma }d_{j\sigma}\rangle=\frac{1}{N}
\sum_{\vec k}{\cal{F}}_{\omega}G_{\vec{k}\sigma}^{(12)}e^{i\vec{k}\cdot \vec{R}_j}$, 
where ${\cal{F}}_{\omega}\Gamma(\omega)\equiv\frac{1}{2\pi i}\oint d\omega f(\omega)\Gamma(\omega)$, in which $f(\omega)$ is the Fermi function 
and $\Gamma(\omega)$ a general Green's function. The Green's functions $G_{\vec{k}\sigma}^{(nm)}$ are obtained as in reference \cite{Calegari1}.
Finally, the correlation functions $\langle S_j^zS_i^z \rangle$ and $h_{ij,-\sigma}$ introduced in section (\ref{model}) are
%
$\langle S_j^zS_i^z \rangle = \frac{(1-b_{ii\sigma})}{2}\left[(n_{-\sigma})^2-h_{ij,-\sigma}^{(1)}\right] 
-\frac{a_{ii\sigma}n_{-\sigma}}{2}$ 
%
and
%
$h_{ij,-\sigma}=\frac{a_{ij,-\sigma}n_{ij,\sigma}+b_{ij,-\sigma}m_{ij,\sigma}}{1+b_{-\sigma}}$
%
with $h_{ij,-\sigma}^{(1)}=\frac{a_{ij,-\sigma}n_{ij,-\sigma}+b_{ij,-\sigma}m_{ij,-\sigma}}{1-b_{ii-\sigma}b_{ii\sigma}}$.

\subsection*{Acknowledgments}
This work was partially supported by the Brazilian agencies CNPq,
CAPES,
FAPERGS
and FAPERJ.
%


\end{document}